\documentclass[a4paper]{panl}
\usepackage{cite}
\usepackage{wrapfig}
\usepackage{graphicx}
\usepackage{amssymb}
\usepackage{amsfonts}
\usepackage{amsmath}
\usepackage{longtable}
\usepackage{rotating}
\usepackage{lscape}
\usepackage{epsfig}
\usepackage{multirow}

\def\veck{\mbox{\boldmath$k$}}

\originalTeX
\begin{document}

\title{Studying the variation of fundamental constants at the Cosmic Ray Extremely Distributed Observatory}
\maketitle
\authors{D.~Alvarez Castillo\/$^{}$\footnote{E-mail: alvarez@theor.jinr.ru}}
\setcounter{footnote}{0}
\from{Institute of Nuclear Physics PAN, Cracow 31-342, Poland}
 
\begin{abstract}
The study of the variation of fundamental constants through time or in localized regions of space is one of the goals of the Cosmic Ray Extremely Distributed Observatory which consists of  multiple detectors over the Earth. In this paper, the various effects which can be potentially identified through cosmic rays detections by CREDO are presented.
\end{abstract}
\vspace*{6pt}

\noindent
PACS: 06.20.Jr; 96.50.S$-$; 04.60.$-$m; 11.30.Cp

\label{sec:intro}
\section*{Introduction}

The purpose of this work is to discuss the possibility of studying the variation of fundamental constants through effects associated to ultra high energy cosmic rays (UHECR) to be detected on Earth. Fundamental constants like the fine structure constant $\alpha$, the gravitational constant $G$ or the speed of light in vacuum $c$ may have had a different value than the one we know, be either in the past or in other places in the Universe, as suggested by diverse theories, like \textit{Quintessence}~\cite{Ratra:1987rm,Uzan:2010pm,Gasperini:2001pc} that attempt to solve cosmological or fundamental physical problems. Under these considerations, cosmic rays might be a probe for variations of those fundamental constants. UHECR produced by a primary particle or a primary photon may extend over large distances resulting in shower footprints on the surface of the Earth of the order of hundreds of kilometers~\cite{Dhital:2018auo,Alvarez-Castillo:2021avz}. The goal of the Cosmic Ray Extremely Distributed Observatory (CREDO)~\cite{CREDO:2020pzy} is to detect those extended showers which might have components correlated in space and/or time. In order to achieve its goal, CREDO consists of a net of detectors mostly in the form of smartphones covering most of the planet. One of the biggest advantage of today's technology is the \textit{multi-messenger approach} that consist of parallel detections in many channels, namely through electromagnetic and gravitational radiation, as well as through cosmic rays. Production of UHECR includes astrophysical mechanisms like Penrose processes in Black Hole emissions~\cite{Stuchlik:2021unj}, evolution and mergers of Neutron Stars\cite{Alvarez-Castillo:2020nkp,Blaschke:2020qqj,Alvarez-Castillo:2019apz}, photon splitting near compact objects~\cite{Usov:2002df}, high energy photons interacting with the magnetic field of the Sun~\cite{Dhital:2018auo,Alvarez-Castillo:2021avz}, dark matter decays~\cite{Garny:2015sjg}, among other scenarios. 

\label{sec:Variation}
\section*{Variation of fundamental constants}

The search for explanation to cosmological observations has resulted in many different conjectures like the existence of dark matter or modifications to Einstein's gravity. As for the latter one, many modifications include the addition of scalar or vector fields that could result in screening mechanisms that would substitute the need for dark matter. Nevertheless, the variation of the gravitational constant $G$ in the past can also provide insights on physics beyond the cosmological paradigm, the so called \textit{Lambda Cold Dark Matter} model ($\Lambda$CDM). For instance, some super string theories suggest a variation of $G$ depending the shape of the potential for the size of the considered internal space scale-lengths in the extra dimensions, additional to the standard four~\cite{Wu:1986ac}. Furthermore, a varying $G$ has been proposed to explain the weakness of the gravitational force with respect to the other fundamental forces, known as the \textit{Dirac large number hypothesis}~\cite{1938RSPSA.165..199D}. One of the most reliable ways to study the time variation of $\dot{G}/G$ is by looking at pulsar radio signals, specially in binary systems where the orbital dynamics strongly depend on $G$. Similarly, highly energetic particles and their associated showers following interaction of the primary source must follow modifications to their constant $G$ expected trajectories. Combined studies from pulsars, cosmic rays and other methods like investigation of stellar and planetary orbits can probe the effects of a changing $G$~\cite{Bisnovatyi-Kogan:2005mjn}.

The variation of the fine structure constant $\alpha$ has been studied through spectral observations of distant quasars, where observations of redshifts $z$ of the components of the fine-structure doublets in their spectra allows to derive $\Delta \alpha/\alpha$ estimates, with typical values of about $\Delta \alpha/\alpha < 10^{-3}$.  When considering the dependence of ultra high energy photons (UHE$\gamma$) and subsequent electromagnetic interactions of their daughter particles after decay, it is clear that $\alpha$ plays a dominant role. The best astrophysical conditions for detections of UHECR on Earth correspond to distances within the so call Greisen–Zatsepin–Kuzmin (GZK) horizon~\cite{Greisen:1966jv,Zatsepin:1966jv} where the particles most likely do not lose energy by interacting with the cosmic microwave background. 

Consideration of the Einstein-Hilbert action together with Quantum Field Theories (QFT) to describe the remaining interactions brings us to two parameters: $G$ and $\Lambda$, the cosmological constant~\cite{Sola:2015xga}. The so called \textit{Planck mass} relates $G$ other fundamental constants: $G=\hbar c /M^{2}_{P}$, where $M^{2}_{P}\approx 1.221 \times 10^{19}$ GeV is the largest mass scale in the Universe. From the dependence of $G$ on $c$ and $\hbar$ it is clear that at energies above $M_{p}c^{2}$ quantum gravity should play a role in the description of the Universe. This hypothesis implies a minimal length, the  \textit{Planck length} $l_{p} = \sqrt{\hbar G /c^{3}} \approx 1.616 \times 10^{-35}$ m, associated to a discrete space-time.

Variation of the speed of light in vacuum is associated with Lorentz Invariance Violation (LIV).  The existence of a minimal length conflicts with Lorentz invariance, because a boosted observer could see the minimal length further Lorentz contracted. In fact, many quantum gravity theories suggest that this phenomenon should occur. The work of \cite{Jacobson:2005bg} considers LIV at high energies, resulting in a modified dispersion relation for a photon.
\begin{equation}	
E_{\gamma}(\veck)=\sqrt{\frac{(1-\kappa)}{(1+\kappa)}}\left|\veck\right|,
\end{equation}
where $\veck$ is the photon momentum and $\kappa$ is a characteristic parameter. Whenever $\kappa>0$ the process of pair production by a UHE$\gamma$ is suppressed, thus more photons could be detected on Earth. The $\kappa=0$ corresponds to the Lorentz Invariance case associated with \textit{normal} pair production. $\kappa<0$ leads to an enhancement of pair-production hindering the detection of  UHE$\gamma$ on Earth. Thus, the critical importance for  UHE$\gamma$ searches. A recently proposed project called \textit{GrailQuest: hunting for Atoms of Space and Time hidden in the wrinkle of Space–Time} aims at monitoring the sky to study high-energy astrophysics phenomena through a swarm of nano/micro/small-satellites to probe the structure of space-time~\cite{GrailQuest:2019ood}. Therefore, a detected delay in photons of different energies in Gamma-ray bursts prompt emissions will allow to constraint or measure the dispersion relation of light in vacuum. Similarly, CREDO which is already in operation, can serve to study photons at the ground level either alone or in connection with different counterparts like gravitational wave detectors. In addition, another quantity depending on fundamental constants is the aforementioned cosmological constant $\Lambda$ which is related to the so called \textit{vacuum energy density} $\rho_{\Lambda}c^{2}=\Lambda c^{4}/(8\pi G)$, the standard candidate for being the Dark Energy of the Universe which causes its accelerated expansion. Its observed value is very low in comparison with the theoretical large value suggested by QFT, a crisis known as the \textit{cosmological constant problem}~\cite{Weinberg:1988cp,Martin:2012bt,Appleby:2018yci}.

The current experimental status of the time variation of constants can be summarized as follows. In 1999 a study based on quasar spectra from the HIRES echelle spectrograph on the Keck I 10m telescope~\cite{Webb:1998cq}, $\Delta \alpha/\alpha$ has been reported to suffer a variation of $10^{-5}$ in the last 10--12 billion years. However, most modern studies with improved accuracy, like observations of absorption systems with the Very Large Telescope coincide with a very slight variation of $10^{-6}$~\cite{Chand:2004ct}. Similarly, the gravitational constant $G$ has been reported to vary not less than $10^{-10}$ per year during the last nine billion years by observations of type Ia supernovae~\cite{Mould:2014iga} under standard Physics assumptions. Interestingly, $G$ has proven difficult to measure with precision resulting in conflicting measurements that suggest a variation of its value~\cite{Anderson:2015bva}. The variation of $c$ is strongly related to the one of $G$ as discussed above. A recent work~\cite{10.1093/mnras/stab1493} based on solar system data instead of cosmological observations takes as a basis the comparison of the unit length defined in two ways: a) defined with a rigid rod or atomic standard, and b) the international standard for the meter based on the transit of light with $c$ as a constant. Therefore, changes in the speed of light in atomic units of length are related to changes in $G$ measured in the two systems. A boundary of $|\dot{c}/c|\leq0.55\times 10^{-12}$yr$^{-1}$ based on experimental constraints of $|\dot{G}/G|$ from spacecraft microwave ranging and from helioseismology has been reported. The authors suggest an extension in epoch relating $c$ to a power scale factor, $a^{n}$. The result, the variation of $c$ being constrained by $|n|< 0.0080$. Contrastingly, the CREDO project seeks to incorporate detections at several scales, from local to cosmological, as well as probing extended cosmic ray detections which in principle should provide complementary constraints to the aforementioned measurements. The magnitudes  of the constraints will correspond to the particular physical scenario being considered. Thus, CREDO represents a new way of studying of physical constants.

\begin{figure}[htpb!]%
 \centering
 \hspace{-1.0cm}\includegraphics[width=0.75\textwidth]{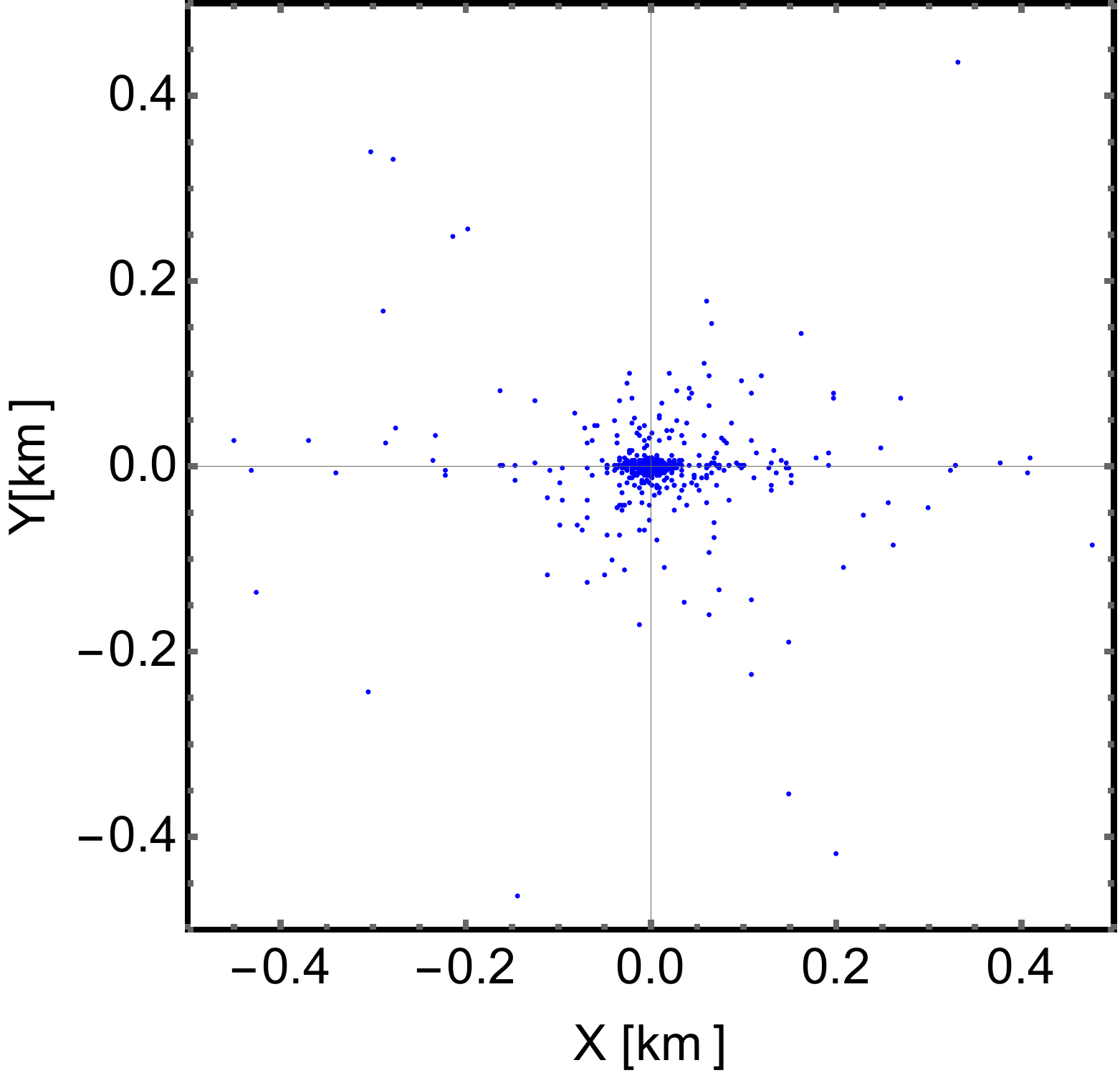}%
\caption{Shower footprint as a signature of an ensemble of cosmic rays. The primary ultra high energy photon (UHE$\gamma$) used in the simulation interacts with the magnetic field of the Sun created an extended cosmic ray shower. Note the characteristic particle distribution with most of the particles lying along the horizontal axis away from the dense core. For details, see~\cite{Dhital:2018auo,Alvarez-Castillo:2021avz}.}%
\label{fig:footprint}%
\end{figure}

Figure~\ref{fig:footprint} shows a simulated particle distribution on Earth at the ground level (footprint) of a Cosmic Ray Ensemble (CRE) produced by a UHE$\gamma$ that is expected to interact with the magnetic field of the Sun. By studying these kind of events through simulations and data searches it will be possible to characterize any deviation from standard scenarios and potentially detect any effect related to LIV or any other phenomena associated with the variation of fundamental physical constants.
In order to identify the relevant detections, extensive data searches by the CREDO collaboration are being currently carried out using novel approaches like Machine Learning techniques~\cite{CREDO:2021yst}.

\section*{Outlook}

Fundamental constants could be varying through time or in particular regions of the Universe. Detection of cosmic rays together with any other of the multi-messenger Astronomy counterparts brings the possibility of testing this hypothesis. In particular, the CREDO experiment is well suited for this task. Some of the main ideas are presented in this manuscript, however quantitative estimations of parameters like distances and energies for efficient detection together with simulations of corresponding CRE footprints are still work in progress. 

\section*{Acknowledgements}

D. A-C. thanks the organizers of the of International Conference on Precision Physics and Fundamental Physical Constants 2021 for their hospitality and attention during the on-site conference in Slovakia. The author acknowledges support from the Bogoliubov Laboratory of theoretical physics in order to attend this conference as well as from the Bogoliubov-Infeld program for collaboration between JINR and Polish Institutions and from the COST actions CA16214 (PHAROS) for networking.

\bibliographystyle{pepan}

\end{document}